\documentclass[aps,prl, superscriptaddress,showpacs,twocolumn]{revtex4}
\usepackage{amsmath,amssymb}

\usepackage{graphicx}

\newcommand{\be}{\begin{eqnarray}}
\newcommand{\ee}{\end{eqnarray}}

\def\ben{\begin{equation}}
\def\een{\end{equation}}
\def\bena{\begin{eqnarray}}
\def\eena{\end{eqnarray}}

\begin{document}

\title{UV-Protected Inflation}

\author{Cristiano Germani}
\email{cristiano.germani@lmu.de}
\affiliation{Arnold Sommerfeld Center, Ludwig-Maximilians-University, Theresienstr. 37, 80333 Muenchen, Germany}

\author{Alex Kehagias}
\email{kehagias@central.ntua.gr}
\affiliation{Physics Division, National Technical University of Athens, 15780 Zografou Campus,  Athens, Greece}




\begin{abstract}
In Natural Inflation, the Inflaton is a pseudo-Nambu-Goldstone boson which acquires a mass by explicit breaking of a global shift symmetry at scale $f$. In this case, for small field values, the potential is flat and stable under radiative corrections. Nevertheless, slow roll conditions enforce $f\gg M_p$ making the validity of the whole scenario questionable. In this letter, we show that a coupling of the Inflaton kinetic term to the Einstein tensor allows $f\ll M_p$ by enhancing the gravitational friction acting on the Inflaton during inflation. This new unique interaction, a) keeps the theory perturbative in the whole inflationary trajectory, b) preserves the tree-level shift invariance of the pseudo-Nambu-Goldstone Boson and c) avoids the introduction of any new degrees of freedom with respect the standard Natural Inflation.

\end{abstract}

\pacs{}

\maketitle
\section{Introduction}

Inflation, a rapid expansion of the early Universe, is a beautiful solution of the homogeneity, isotropy and flatness puzzles \cite{mukbook}. 
Although an isotropic expansion of the Universe might be obtained by considering general non-minimally coupled p-forms \cite{muk}, the minimally coupled zero-form (a scalar field) is the most simple and natural source for inflation \cite{inf,linde,stei}. In this case, the scalar field during inflation generates an almost DeSitter (exponential) expansion of the early Universe. 

In a Friedmann-Robertson-Walker spacetime (FRW) with metric $ds^2=-dt^2+a(t)^2d\vec{x}\cdot d\vec{x}$, a minimally coupled scalar field ($\phi$) to gravity, with a potential $V(\phi)>0$, produces an acceleration $\ddot a\sim -(\dot\phi^2-V)$, where $(\dot{})=d/dt$.
It is then clear that to get an accelerated expansion ($\ddot a>0$) for a long time, the field $\phi$ has to ``slow roll'' in its own potential, i.e. $\dot \phi^2\ll V$. Unfortunately though, while solving the cosmological puzzles, this seemly innocuous condition threaten the whole inflationary paradigm, as we shall see.

During slow roll, the Hubble equation and field equations for the Inflaton are
\be\label{eq0}
H^2\simeq \frac{V}{3M_p^2}\ ,~~~~\dot\phi\simeq -\frac{V'}{3 H}\ ,
\ee
where $H=\dot a/a$ is the Hubble ``constant'',  $(')=d/d\phi$ and $M_p$ is the Planck scale. Eqs. (\ref{eq0}) are found by considering the slow roll conditions
\be\label{slow}
\epsilon\equiv -\frac{\dot H}{H^2}\ll 1\ ,~~~~\delta\equiv \Big|\frac{\ddot\phi}{3H\dot\phi}\Big|\ll 1\ .
\ee
By using (\ref{eq0}) and pluging into (\ref{slow}) we find the two independent conditions
\be\nonumber
\epsilon\simeq M_p^2\frac{V'^2}{2V^2}\ll 1\ ,~~~~\eta\simeq \Big| M_p^2\frac{V''}{3V}\Big|\ll 1\ .
\ee

A common problem of standard inflationary scenarios is the ``eta'' problem. Gravity is not a renormalizable theory as its perturbative expansion breaks down at the Planck scale $M_p$. Therefore, operators suppressed by the scale $M_p$, although not known, are generically expected to complete the theory at UV. In particular, one would expect the Inflaton potential to be generically ``corrected'' by higher dimensional operators. Consider for example the following dimension six correction $\tilde V=V\times (1+c\frac{\phi^2}{M_p^2}+\ldots)$,
where $c$ is an unknown coefficient expected to be of ${\cal O}(1)$. With this correction, we have that the $\eta$ parameter is modified as
$\tilde \eta\simeq \eta+c\frac{2}{3}+\ldots$,
therefore, if $\eta\ll 1$, in order to keep $\tilde\eta\ll 1$ we need $c\ll 1$. However, this coefficient cannot be calculated unless the full UV completed theory of gravity is known. Note that for small field scenarios this is the leading correction to $\eta$. In large field scenario the problem is clearly far more severe as an infinite amount of non-renormalizable operators has to be set exactly to zero.

The ``eta'' problem might be nevertheless naturally solved in small fields scenarios by introducing new symmetries in the Inflaton lagrangian. These symmetries might in fact force the potential to be flat even under radiative corrections. 
There are two possible symmetries to achieve this goal: global and local. 

Local symmetries (such as local supersymmerty) might stabilize the Inflaton potential in supergravity. Still, in this framework, supersymmetry is explicitly broken by the gravitational background making the Inflaton potential generically too steep to produce inflation \cite{Lyth:1998xn}. Additional local symmetries related to the matter content of the full theory in which the Inflaton is embedded might nevertheless change this no-go result, as proposed in \cite{bau}. However, to date, this direction is still under development.

The other possibility for a radiative stability of the Inflaton potential is the existence of global symmetries. A use of global symmetries in inflation is encoded in the so called Natural Inflation \cite{natural}. 
In Natural Inflation, the Inflaton is a pseudo-Nambu-Goldstone boson acquiring a mass by explicit breaking of a global shift symmetry at scale $f$. This happens by instanton effects, as in the Peccei-Quinn mechanism \cite{pq}. In this case, for small field values, the potential is flat and stable under radiative corrections. Nevertheless, the slow roll condition $\eta\ll 1$ implies $f\gg M_p$ making the whole scenario unreliable.

In the following, we will show that the scale $f$ might be safely taken to be much smaller than the Planck scale by introducing a non-minimal coupling of the Inflaton kinetic term to the Einstein tensor. In particular, we will show that this new theory is in the weak coupling regime during the whole inflationary evolution and does not propagate any more degrees of freedom than the ones already existing in Natural Inflation.

\section{Constructing the action}

We now consider the most generic theory such that

\begin{itemize}
\item It is diffeomorphisms invariant;
\item It only propagates a (non-ghost) massless spin $2$ and a spin $0$ particle on any background;
\item It is tree-level shift symmetric in the field $\phi$ (i.e. is symmetric under $\phi\rightarrow \phi+c$).
\end{itemize}

Such action has been found in \cite{new} and it is
\begin{eqnarray}\label{action0}
{\cal S}=\frac{1}{2}\int d^4x\sqrt{-g}\left[M_p^2 R-\Delta^{\alpha\beta}\partial_\alpha\phi\partial_\beta\phi\right] \ .
\end{eqnarray}
In (\ref{action0}) we used the notation
\be\label{delta0}
\Delta^{\alpha\beta}=g^{\alpha\beta}-\frac{1}{M^2} G^{\alpha\beta}\ ,
\ee
where $G^{\alpha\beta}\equiv R^{\alpha\beta}-\frac{1}{2}Rg^{\alpha\beta}$ and $M$ are respectively the Einstein tensor and a new mass scale, not necessarily related to the Planck mass. Note that the minus sign in the
definition (\ref{delta0}) is crucial to avoid ghosts, whenever weak energy conditions are satisfied \cite{wald}. One may wonder whether an infinite series of curvatures non-minimally coupled to the kinetic term of the scalar, can be added to (\ref{action0}). We claim that it is unlikely that a fine tuning exist in which  {\it both} metric and scalar equation of motions are second order in derivatives.  

In Natural Inflation the massless field $\theta$ is a pseudo-Nambu-Goldstone field with decay constant $f$ and periodicity $2\pi$.
Let us consider the following tree-level lagrangian for $\theta$
\be\label{axion}
{\cal S}=\int d^4x\sqrt{-g}\Big[\frac{M_p^2}{2} R-\frac{f^2}{2}\Delta^{\alpha\beta}\partial_\alpha
\theta\partial_\beta \theta+\cr +me^{i\theta}\bar\psi(1+\gamma_5)\psi+\bar\psi{\not} {\cal D} \psi-\frac{1}{2}\mbox{Tr}F_{\alpha\beta}F^{\alpha\beta}\Big] \ ,
\ee
where $\psi$ is a fermion charged under the (non-abelian) gauge field with field strength $F_{\alpha\beta}$, ${\not} {\cal D} =\gamma^\alpha{\cal D} _\alpha$ is the gauge invariant derivative and $m$ is a mass scale.

The action (\ref{axion}) is invariant under the chiral (global) symmetry $\psi\rightarrow e^{i\gamma_5\alpha/2}\psi$, where $\alpha$ is a constant. This symmetry is related to the invariance under shift symmetry of $\theta$, i.e. $\theta\rightarrow\theta-\alpha$.

The chiral symmetry of the system is however broken at one loop level \cite{peskin} giving the effective interaction $\theta F\cdot \tilde F$, where $\tilde F^{\mu\nu}=\sqrt{-g}\epsilon^{\alpha\beta\mu\nu}F_{\alpha\beta}$ and $\epsilon^{\alpha\beta\mu\nu}$ is Levi-Civita antisymmetric symbol. Instanton effects related to the gauge theory $F$ introduce a potential $K(F\cdot\tilde F)$ \cite{dvali2}. In the zero momentum limit we can integrate out the combination $F\cdot\tilde F$ and obtain a periodic potential for the field $\theta$ (note that this is independent upon the canonical normalization of $\theta$) which has a stable minimum at $\theta=0$ \cite{vafa}. If we expand the potential around its own maximum at $\theta=\theta_{\rm max}$, we get
 \be\label{pot}
 V(\phi)\simeq\Lambda^4\left(1-\frac{\phi^2}{2f^2}\right)\ ,
 \ee
 where $\Lambda$ is the strong coupling scale of the gauge theory $F$ \cite{gross} and $\phi=f(\theta-\theta_{\rm max})$. The approximation (\ref{pot}) is valid as long as $\phi\ll f$ and it is precisely in this regime that the Universe can naturally inflate, as it is shown later on.
  
What is very important to note is that since the non-linearly realized symmetry is restored as $\Lambda\rightarrow 0$, 
the potential (\ref{pot}) is natural and no UV corrections may spoil its flatness.

Finally, the action we will use for the inflationary background is therefore
\be\label{action}
{\cal S}=\int d^4x\sqrt{-g}\Big[\frac{M_p^2}{2} R-\frac{1}{2}\Delta^{\alpha\beta}\partial_\alpha
\phi\partial_\beta \phi-V(\phi)\Big] \ .
\ee

\section{Strong coupling scale}

In order to find the validity range of the effective action (\ref{action}) we should find the energy scale in which non-renormalizable operators become strong. Obviously, in a non-minimal coupled model, this is a background dependent question as already noted in \cite{new, shap2}. In a non-trivial background for the scalar field configuration one might employ the gauge $\delta\phi=0$ (at all orders), which is quite useful in cosmological perturbations theory \cite{malda}. In this case the metric is perturbed as $g_{\alpha\beta}=\gamma_{\alpha\beta}+\frac{h_{\alpha\beta}}{M_p}$, where $\gamma_{\alpha\beta}$ and $h_{\alpha\beta}$ are respectively the background and the perturbed metrics. We now use the ADM formalism where the metric is generically written as $ds^2=-N^2dt^2+h_{ij}\left(dx^i-N^idt\right)\left(dx^j-N^jdt\right)$.
In this form the degrees of freedom of the graviton are encoded into $h_{ij}$ and $N,N^i$ are lagrange multiplier in the action (\ref{action}) \cite{new}. In this formalism one might define the extrinsic curvature $K_{ij}=\frac{1}{N}\left(\dot h_{ij}-D_{(i} N_{j)}\right)$ and the three curvature $\cal R$ where the covariant derivative $D_i$ and $\cal R$ are both calculated with the three-metric $h^{ij}$ \cite{wald}. The perturbed action (\ref{action}) is then \cite{new2}
\be\label{delta}
S_\delta=\frac{1}{2}\int d^3x dt\sqrt{h}N\Big (M_p^2{\cal R}(1+\frac{\dot{\phi}^2}{2M^2 M_p^2})+\cr
+M_p^2(K_{ij}K^{ij}-K^2)(1-\frac{\dot{\phi}^2}{2M^2 M_p^2})+
\frac{\dot{\phi}^2}{N^2}-
2V\Big)\ .
\ee
As we shall see, during slow roll inflation, $\frac{\dot\phi^2}{M^2 M_p^2}\ll 1$ and considered roughly constant. Therefore (\ref{delta}) is well approximated as
\be\label{sc}
S_\delta\simeq\frac{1}{2}\int d^4x \sqrt{-g} \Big (M_p^2 R+N^{-2}
\dot{\phi}^2-2V\Big)\ ,
\ee
so that the strong coupling scale of (\ref{sc}) is manifestly $M_p$. Note that in the case of multi-fields coupled to (\ref{delta0}), as in the New Higgs Inflation \cite{new}, the strong coupling scale is lower \cite{new}.

Another source for a strong coupling scale is the operator $S_{int}=\int d^4x \sqrt{-g}\frac{\phi}{f}F\cdot\tilde F$, which was integrated out to get the action (\ref{action}) \footnote{CG thanks Fedor Bezrukov for pointing this out and Yuki Watanabe for discussions.}. 

In ADM formalism, and during slow roll, the scalar perturbations of the metric are codified into the canonically normalized scalar perturbation $\bar\zeta$ in $h_{ij}=a(t)^3(1+2\frac{M}{\dot\phi}\bar\zeta)\delta_{ij}$ \cite{new2}. After integrating by parts $S_{int}$ in the $\delta\phi=0$ gauge, we get the perturbed action
\be
\delta S_{int}\sim\int d^4x \sqrt{-\gamma} \frac{M}{f}\delta C_{ijk}\epsilon^{0ijk}\left(\bar\zeta+\frac{\dot{\bar\zeta}}{H}\right)\ ,\nonumber
\ee
where $\delta C_{\alpha\beta\gamma}$ is the perturbed Chern-Simons three-form relater to the gauge field $F$ \cite{dvali2}. This interaction has a renormalizable and a non-renormalizable term. The renormalizable term is always in the weak regime as long as $f\gg M$. The non-renormalizable interaction however, introduce a new strong coupling scale $M_{new}=H\frac{f}{M}$. The strong coupling scale of the theory will then be $M_*={\rm min}\left(M_p, M_{new}\right)$. Note that in the Minkowski limit the canonical normalization of $\bar\zeta$ is different and it boils down into the replacement of $H$ instead of $M$ in $M_{new}$, so that $M_{new}\rightarrow f$, as expected. Moreover, because slow roll conditions are violated in the Minkowski limit ($M_p H\sim\dot\phi\rightarrow 0$), the unitary violating scale related to the operator $\Delta_{\alpha\beta}$ smoothly converge to $M_*\sim (M_p M^2)^{1/3}$. What is important to note is that during the whole evolution, from the Inflationary to the Minkowski background, the curvature is always below the scale $M_*$, so that our theory can be perturbatively trusted. 

\section{UV-protected Inflation}

The variation of the action (\ref{action}) with respect to the lapse $N$ and the field $\phi$ give rise to the two Hubble and field equations \cite{new}
\be
H^2=\frac{1}{6 M_p^2}\left[\dot\phi^2\left(1+9\frac{H^2}{M^2}\right)+2 V\right]\ ,\cr\ 
\partial_t\left[a^3\dot\phi\left(1+3\frac{H^2}{M^2}\right)\right]=-a^3  V'\ .\label{background}
\ee
We will ask the solution to obey the following inequalities
\be\label{slowroll}
H^2\gg M^2\ , ~~~  \epsilon\equiv-\frac{\dot H}{H^2}\ll1\ , ~~~ \delta\equiv \Big|\frac{\ddot\Phi}{3H \dot\Phi}\Big|\ll 1\ ,
\ee
where the last two are the usual slow roll conditions. In particular the second implies that during inflation the universe undergoes an exponential expansion.

With the help of equations (\ref{background}) we find the following independent conditions extracted from (\ref{slowroll}):
\be\label{slowroll23}
\eta\simeq\frac{M^2}{9H^2}\frac{M_p^2}{f^2}\ll1\ ,\ \epsilon\simeq\frac{1}{6}\frac{M^2}{H^2}\frac{\phi^2}{f^2}\frac{M_p^2}{f^2}\ll 1\ ,\ H^2\gg M^2\ .
\ee
Note that both $\eta$ and $\epsilon$ are suppressed by the additional gravitational friction term $\frac{H^2}{M^2}\gg 1$ which is not present in the standard Natural Inflation \cite{natural}. This enhanced gravitational friction is the key physical mechanism allowing $f\ll M_*$. 

Combining the weak coupling constraint ($f\gg M$) and the $\eta$ constraint in (\ref{slowroll23}), we find $M_{new}\gg M_p$. Therefore, during slow roll, $M_*\simeq M_p$. The Quantum Gravity constraint such that the curvature should be smaller than the Planck scale, is easily satisfied for $\Lambda^4\ll M_p^4$. The friction constraint $H^2\gg M^2$ is satisfied for $\Lambda^4\gg M^2 M_p^2$ which implies $M^2\ll M_p^2$ as it should. Finally we would like to impose $f\ll M_p$.

Collecting all constraints, the Natural Inflationary set-up is UV protected if the following hierarchies of scales are satisfied
\be\label{conditions}
M_p^4\gg\Lambda^4\gg M^2 M_p^2\ ,~~~~\frac{M M_p}{\Lambda^2}\ll\frac{f}{M_p}\ll 1\ .
\ee

\section{Quantum gravity corrections} 

In this section we will address potential issues related to quantum (gravity) corrections to our inflationary scenario.

For any theory of gravity which does not propagate ghosts and it approaches General Relativity at large distances/weak curvatures, the strong coupling scale $M_*$ of the theory might only be below the Planck scale $M_p$ \cite{dvali}. We showed that, during inflation, the strong coupling scale of our set-up is indeed below $M_p$. In this respect, our theory does not propagate any hidden ghost. 

A second issue is related to the fact that, in any healthy field theory, one expects that all scales playing any role in the weak regime, are smaller than the strong coupling scale of the theory. This requirement is obviously fulfilled by (\ref{conditions}) as $\Lambda, f,M\ll M_*$. In the absence of the gravitational enhanced friction (or the operator (\ref{delta0})), one necessarily needs $f\gg M_*$ in order to produce inflation. In this respect then, Natural Inflation \cite{natural} is problematic.
  
Another issue is related to the global symmetry of the tree-level action (\ref{axion}). It is widely believed that Quantum Gravity does not allow global symmetries. This is due to the fact that Black Hole evaporation democratically emit any particle coupled to gravity. The only constraint on this emission is to conserve the total energy and/or fluxes at infinity. Global charges do not carry any flux and therefore cannot be conserved.
In our scenario, however, the global (shift) symmetry is already broken by gauge instanton effects. Therefore, the only way that gravity may participate to the quantum breaking of the global chiral symmetry of (\ref{axion}) is through gravitational anomalies. The latter, can only couple to the axion via the interaction $\theta R\tilde R=\theta \sqrt{-g}\epsilon^{\alpha\beta\mu\nu}R^\gamma{}_{\delta\alpha\beta}R^\delta{}_{\gamma\mu\nu}$, related to the gravitational Chern-Simons three-form \cite{dvali2}. A potential $K_G(R\tilde R)$ might then be generated by instanton effects if and only if the zero momentum limit of the instanton correlator  $\langle R\tilde R,R\tilde R\rangle$ does not vanish \cite{dvali2}. In any case, even if the potential $K_G(R\tilde R)$ is generated, it is suppressed by the factor $e^{-S}\ll 1$, where $S$ is the instanton action. Quantum gravity effects can therefore only produce a small correction to the mass of the Inflaton field by redefining a new effective $\Lambda$. 

Although the potential we introduced is stable under radiative corrections, one might wonder about derivative terms generated by loops corrections. During slow roll inflation, all these corrections are negligible as proportional to the slow roll parameters and suppressed by the scale $M_*$.

Concluding, in our case in which $f,\Lambda, M, H\ll M_*$, there are no substantial quantum (gravity) corrections to the inflationary evolution.

\section{Conclusions}
A pseudo-Nambu-Goldstone scalar (the axion), whose potential is obtained by a global symmetry breaking at scale $f$ via gauge field instanton effects, has a naturally flat potential, as long as $f\ll M_p$. Unfortunately though, slow roll conditions for the axion require $f\gg M_p$. This requirement might be softened by introducing a plethora of $N\gg 1$ axions as in \cite{n}, if large cross interaction among the fields can be avoided. In this case, the effective friction acting on the radial direction in the field space, is boosted by a factor $N$ so that slow roll conditions require only $f\gg M_p/\sqrt{N}$. However, in this framework, the strong coupling scale of the theory is lowered to $M_*\sim M_p/\sqrt{N}$ \cite{giabh}. This nullifies the attempt of \cite{n} to produce a natural inflationary scenario \cite{huang}. This problem may however be solved if only two axions are considered. In this case, by a fine adjustment of their coupling to the anomalous currents, one can find $f\ll M_*$ \cite{peloso}. 

In this paper we showed an alternative way to increase the friction in a Natural Inflationary scenario without introducing any new degrees of freedom. In our model the friction of the axion is gravitationally enhanced. In this case, in order for the axion to slow roll down its own potential, a natural value for the global symmetry breaking scale $f\ll M_p$ (or more precisely $f\ll M_*$) can be easily obtained. This feature is uniquely obtained by the interaction of the Einstein tensor to the kinetic term of the axion which keeps nevertheless the theory perturbative during the whole inflationary evolution. This interaction is unique in the sense that it does not propagate more degrees of freedom than a massless spin 2 and a scalar, while keeping the tree-level shift invariance of the axion untouched.\\
{\it Acknowledgments} CG thanks Fedor Bezrukov, Savas Dimopoulos, Gia Dvali, Viatcheslav Mukhanov, Alex Pritzel, Eva Silverstein and Yuki Watanabe for important discussions and comments. CG is sponsored by the Humboldt Foundation. This work is partially supported by the PEVE-NTUA-2009 program.

\end{document}